\documentclass[10pt]{article}
\usepackage{graphicx}

\begin{document}

 \twocolumn [ \noindent {\footnotesize\it  \small ISSN 1063-7737,
 Astronomy Letters, 2008, Vol. 34, No. 8, pp. 509--514.
 \copyright Pleiades Publishing, Inc., 2008.

\noindent \small Original Russian Text \copyright V.~B.~Petkov,
E.~V.~Bugaev, P.~A.~Klimai, M.~V.~Andreev, V.~I.~Volchenko,
G.~V.~Volchenko, A.~N.~Gaponenko, Zh.~Sh.~Guliev, I.~M.~Dzaparova,
D.~V.~Smirnov, A.~V.~Sergeev, A.~B.~Chernyaev, A.~F.~Yanin, 2008,
published in Pis'ma v Astronomicheski$\check{\imath}$ Zhurnal,
2008, Vol. 34, No. 8, pp. 563--569.}

 \vskip -4mm

\begin{tabular}{llllllllllllllllllllllllllllllllllllllllllllllll}
 & & & & & & & & & & & & & & & & & & & & & & & & & & & & & & & & & & & & & & & \\
\hline \hline
\end{tabular}

\vskip 1.5cm

 \centerline {\Large\bf Searching for Very-High-Energy Gamma-Ray Bursts from }
 \centerline {\Large\bf Evaporating Primordial Black Holes}
 \bigskip
 \centerline {\large\bf V.B. Petkov$^1$, E.V.
Bugaev$^1$, P.A. Klimai$^1$, M.V. Andreev$^2$, V.I. Volchenko$^1$,} %
\centerline {\large\bf G.V. Volchenko$^1$, A.N. Gaponenko$^1$,
Zh.Sh.
Guliev$^1$, I.M. Dzaparova$^1$, } %
\centerline {\large\bf D.V. Smirnov$^1$, A.V. Sergeev$^2$, A.B.
Chernyaev$^1$, A.F. Yanin$^1$}

 \medskip
 \centerline {\it $^1$ Institute for Nuclear Research, Russian Academy of Sciences,}
 \centerline {\it ul. 60-letiya Oktyabrya 7a, Moscow, 117312 Russia}

 \medskip
 \centerline {\it $^2$ International Center for Astronomical and Medicoecological Research,}
 \centerline {\it National Academy of Sciences of Ukraine, Ukraine}

 \medskip
 \centerline {\small Received December 6, 2007}

 \medskip
 \centerline {\small E-mail: {\tt pklimai@gmail.com}}

 \bigskip

{\noindent {\bf Abstract---}\small Temporal and energy
characteristics of the very-high-energy gamma-ray bursts from
evaporating primordial black holes have been calculated by
assuming that the photospheric and chromospheric effects are
negligible. The technique of searching for such bursts on shower
arrays is described. We show that the burst time profile and the
array dead time should be taken into account to interpret
experimental data. Based on data from the Andyrchy array of the
Baksan Neutrino Observatory (Institute for Nuclear Research,
Russian Academy of Sciences), we have obtained an upper limit on
the number density of evaporating primordial black holes in a
local region of space with a scale size of $\sim 10^{-3}$ pc.
Comparison with the results of previous experiments is made. }
\bigskip


{\bf DOI}: 10.1134/S106377370808001X


\vskip 1cm

]

\section*{Introduction}

Primordial black holes (PBHs) can be formed in the early Universe
through the gravitational collapse of primordial cosmological
density fluctuations. Therefore, the formation probability of PBHs
and their observational manifestations depend significantly on how
the primordial density fluctuations emerged and developed. The
pattern of black hole formation is determined not only by the
cosmology and physics of the early Universe. Theoretical
predictions of the PBH formation probability depend strongly on
the adopted theory of gravitation and, which is also important, on
the adopted model of gravitational collapse. The evaporation of
black holes on which their experimental search is based has not
been completely studied either. Thus, PBH detection will provide
valuable information about the early Universe and can be a unique
test of the general theory of relativity, cosmology, and quantum
gravity (Carr 2003). Knowledge of the spatial distribution of PBHs
is important for their direct search. As was shown by Chisholm
(2006), the local PBH number density in our Galaxy could be many
orders of magnitude higher than the mean PBH number density in the
Universe (the density ratio could reach $\sim 10^{22}$, which is
much larger than the previously predicted value of $\sim 10^7$).
Therefore, the constraints on the PBH number density imposed by
direct searches can be more stringent than those imposed by
diffuse extragalactic gamma-ray background measurements.

The high-energy gamma-ray bursts (GRBs), i.e., significant and
time-localized excesses of gamma radiation above the background,
are generated at the final PBH evaporation stage. Since the
calculated temporal and energy characteristics of such bursts
depend on the theoretical evaporation model (Bugaev et al. 2007),
the technique of an experimental search and, accordingly, the
constraints imposed on the PBH number density in the local
Universe are modeldependent. PBHs can be searched for in
experiments on arrays designed to detect extensive air showers
(EASs) from cosmic rays with effective primary gamma-ray energies
of 10 TeV or higher only within the framework of the evaporation
model without a chromosphere (MacGibbon and Webber 1990). The
duration of the high-energy GRBs predicted by chromospheric
evaporation models is too short, much shorter than the dead time
of EAS arrays.

It should be noted that the duration of the highenergy GRBs is
fairly short in the evaporation model without a chromosphere as
well. Therefore, the effect of the array dead time on the burst
detection probability should be taken into account when
interpreting the experimental data from EAS arrays with a high
threshold energy of the primary gamma-ray photons.

\begin{figure}[t]
{
\includegraphics[width=82mm]{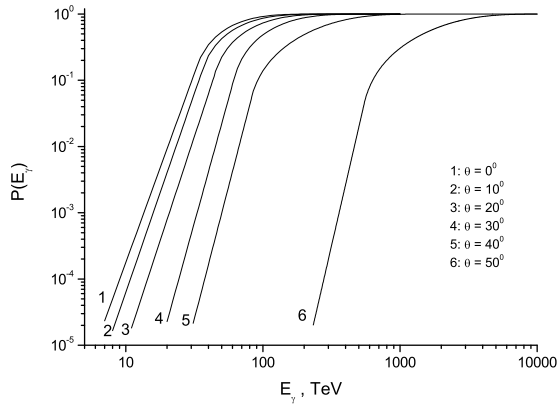}
\caption{ Probability of EAS detection by the Andyrchy array
versus primary gamma-ray photon energy for various zenith angles
$\theta$. } }
\end{figure}

\section*{The Experiment}

The Andyrchy array to detect EASs from cosmic rays is located on
the flank of Mount Andyrchy at an altitude of $\sim 2060$ m above
sea level; its geographical coordinates are $43.28^\circ$ N and
$42.69^\circ$ E. The array consists of 37 scintillation detectors
based on plastic scintillator, each with an area of 1 m$^2$. The
separation between the detectors in the horizontal plane is $\sim
40$ m and the total area of the array is $5 \times 10^4$ m$^2$.
The EAS trigger becomes active when $\ge 4$ array detectors are
triggered simultaneously; the trigger rate is $\sim 9$ s$^{-1}$.
The effective angular resolution of the array for such events is
$3.8^\circ$. The array dead time per EAS event does not depend on
the EAS power and is 1 ms. The array and its operating parameters
were described in detail previously (Petkov et al. 2006).

The detection probabilities $P(E_\gamma, \theta)$ of the EASs
generated by primary gamma-ray photons with energy $E_\gamma$
incident on the array at zenith angle $\theta$ were determined by
simulating electromagnetic cascades in the atmosphere and the
detector (Smirnov et al. 2005; Smirnov 2005). The CORSIKA code
(version 6.03, the QGSJET model) (Heck et al. 1998) was used to
simulate electromagnetic cascades in the atmosphere. The
characteristics of the secondary particles that reached the array
level were used as input parameters in the code for calculating
the detector response, in which the energy release and the
triggering time of each array detector were calculated. For the
event simulated in this way, the arrival direction of the
simulated EAS was reconstructed on the basis of a standard
technique used in processing the recorded events. In Fig. 1, the
EAS detection probability is plotted against the primary gamma-ray
photon energy for several zenith angles. Since the detection
probability of primary gamma-ray photons is a relatively smooth
function of the photon energy, the median energy of the primary
gamma-ray photons detected by the array depends on their energy
spectrum. Following Alexandreas et al. (1993), we will take the
median energy of the primary gammaray photons when the source is
located at zenith and the gamma-ray spectrumis a power law with an
index of $-2.7$ as the effective energy of the gamma-ray photons
detected by the array. For the EASs detected by the Andyrchy
array, this energy is 60 TeV.

For each of the events selected by the EAS trigger, we
reconstructed the EAS arrival direction, i.e., the zenith and
azimuth angles $(\theta, \phi)$ in the array coordinate system.
Based on our processing, we created an archive of preprocessed
information for the period 1996–-2001 (the net accumulation time
is $\sim1100$ days and the total number of events is $\sim6.22
\times 10^8$), which contains the absolute event time (with an
accuracy of 1 ms) and the EAS arrival direction. Searching for
GRBs over the celestial sphere (without referencing to the already
detected bursts) is, in fact, searching for spatiotemporal
concentrations of events (clusters). Since we take fairly short
time intervals, spatial concentrations of events are searched for
in the horizontal coordinate system. For each event $i$ with an
absolute time $t_i$ and arrival angles $(\theta, \phi)_i$, we
search for a cluster of such events $i, i+1, ... , i+n-1$ that the
EAS arrival directions differ by less than $\alpha_r$ from the
weighted mean direction. Thus, each cluster is characterized by
multiplicity $n$, duration $\Delta t$, absolute time $T$, and
arrival direction $(\theta, \phi)$.

Previously, data from the Andyrchy array (Smirnov et al. 2005;
Smirnov 2005) were used to search for cosmic GRBs over the
celestial sphere. Groups of EASs arrived from one angular cell
$\alpha_r = 4.0^\circ$ in radius were selected; the minimum and
maximum time differences in the cluster were taken to be 10 ms and
10 s, respectively. For each multiplicity ($n \ge 2$), the
dependences of the number of clusters with a given multiplicity on
the interval $\Delta t$ were derived. The background of chance
coincidences (the formation of clusters with a given multiplicity
$n$) was calculated using a similar processing of the simulated
events. The EAS arrival angles $(\theta, \phi)$ and the time
between the EAS arrivals were simulated using experimental
distributions. The experimentally measured detection rates of such
clusters are in agreement with those expected from the background
of chance coincidences (within the limits of one standard
deviation).

\section*{Searching for GRBs from PBHs for the Evaporation Model
Without a Chromosphere}

The spectrum of the gamma-ray photons emitted by PBHs,
$dN_\gamma/dE_\gamma$, depends on the time $t_l$ left until the
end of black hole evaporation. In the evaporation model without a
chromosphere (MacGibbon and Webber 1990), the numerically
calculated and time-integrated photon spectrum can be fitted by a
piecewise power function (for a black hole temperature $T_H \gg
m_\pi$ and gamma-ray energies $E_\gamma > m_\pi/2$):
\begin{eqnarray}
 \frac{dN_{\gamma}}{dE_{\gamma}} = N_0 \left\{ \begin{array}
{l}
  \Big( \frac{E_0}{T_H} \Big) ^3 \Big(\frac{T_H}{E_{\gamma}}\Big)^{3/2} \;\; , \;\; E_{\gamma} < T_H \\
  \Big( \frac{E_0}{E_{\gamma}} \Big)^3 \;\;\;\; , \;\;\;\; E_{\gamma} \ge T_H
 \end{array} \right.
  \nonumber
\end{eqnarray}
(Bugaev et al. 2007), where $E_0 = 10^5$ and all energies are
measured in GeV. The parameter $N_0$ is
$$ N_0 \approx \nu_0 \times 7\cdot 10^{19} , $$
while $\nu_0$ includes the effective number of degrees of freedom
of the quarks and gluons used in calculating the photon spectrum.
A simple estimate for $N_0$ includes the $u$ and $d$ quarks, their
antiquarks, and all gluons; in this case, $\nu_0 = (1/3)(3 \cdot 2
\cdot 2 \cdot 2 + 8 \cdot 2) = 40/3$ ($1/3$ is the ratio of the
number of $\pi^0$ mesons to the total number of $\pi$ mesons). The
corresponding value of $N_0$ is $9 \times 10^{20}$. The lower
limit for the parameter $\nu_0$ can be obtained by taking into
account only one type of quarks ($\nu_0 = 4$); in this case, $N_0
\approx 3 \times 10^{20}$.

The time until the end of PBH evaporation $t_l$ (in seconds) is
related to the black hole temperature $T_H$ (in GeV) by
$$t_l = 4.7 \times 10^{11} \; T_H^{-3} .$$

Since the array detects the EASs generated by primary gamma-ray
photons with energy $E_\gamma$ incident at zenith angle $\theta$
with probability $P(E_\gamma, \theta)$ and since the spectrum of
the gamma-ray photons emitted by PBHs $dN_\gamma/dE_\gamma$
depends on $t_l$, the spectrum recorded by the array (its response
function) $P(E_\gamma, \theta) dN_\gamma/dE_\gamma$ also depends
on $t_l$. Here, $\theta$ is the zenith angle at which the PBH is
seen from the array. Figure 2 shows the photon spectra and
response functions of the Andyrchy array for the vertical
direction and for $t_l = 1$ s and $1$ ms, with $N_0 = 3\times
10^{20}$.

\begin{figure}[!t]
{
\includegraphics[width=82mm]{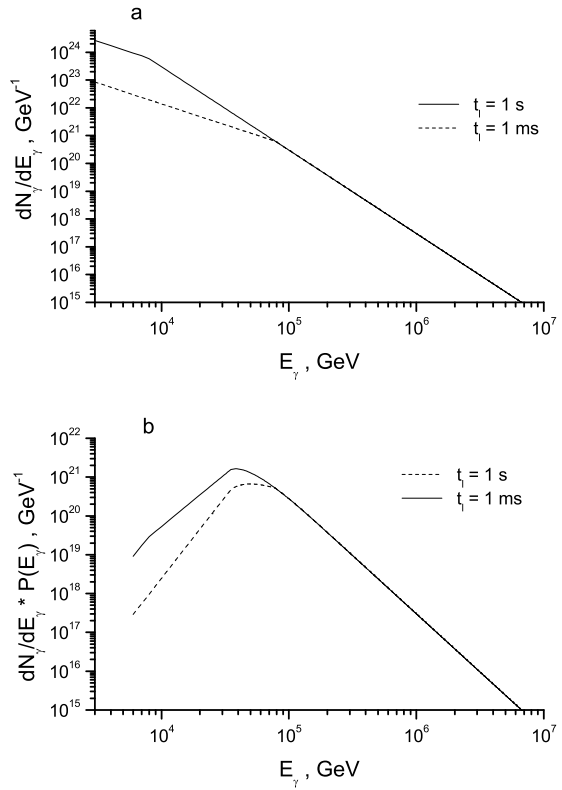}
\caption{ (a) Photon spectrum from evaporating PBHs. (b) Response
functions of the Andyrchy array for the vertical direction at two
values of $t_l$. } }
\end{figure}

The total number of gamma-ray photons emitted by PBHs that can be
detected by the array,
$$N_{\gamma}(\theta, t_l)= \int\limits_{0}^{\infty}
dE_\gamma  P(E_{\gamma},{\theta})  dN_{\gamma}/dE_{\gamma} ,$$
also depends on $t_l$. Figure 3 shows the dependences
$N_\gamma(\theta, t_l)$ (integrated burst profile) for several
zenith angles $\theta$. It should be noted that the burst profile
mainly determines the technique of searching for GRBs from PBHs on
a specific array, in particular, the choice of a time interval
$\Delta t$ for the search for EAS clusters. Let us call the time
left until the end of black hole evaporation during which $99\%$
of the gamma-ray photons that can be detected by a given array are
emitted the burst duration $t_b$ for this array. In Fig. 4, $t_b$
is plotted against the zenith angle for the Andyrchy array. We see
from the figure that the array can record the events from
evaporating PBHs in a limited range of zenith angles, because the
expected GRB duration at large zenith angles is shorter than the
array dead time per event $t_d$.

We searched for the groups of EASs arrived from one angular cell
for time intervals of a given duration, $\Delta t = t_b(\theta)$.
Because of the short time intervals (accordingly, the reduced
background of chance coincidences), we used large angular cells:
$\alpha_r = 7.0^\circ$; such a cell contains $90\%$ of the events
from a point source. The distributions of the detected clusters in
multiplicity were obtained for $\Delta t = 40, \;24, \;11,$ and
$3$ ms, which correspond to the zenith angles $\theta = 0^\circ,
10^\circ, 20^\circ,$ and $30^\circ$. The results of our search for
two values of $t_b$ are presented in Fig. 5. The maximum
multiplicities $n'(\theta)$ of the detected clusters are 4, 4, 3,
and 2, respectively.

\begin{figure}[!t]
{
\includegraphics[width=82mm]{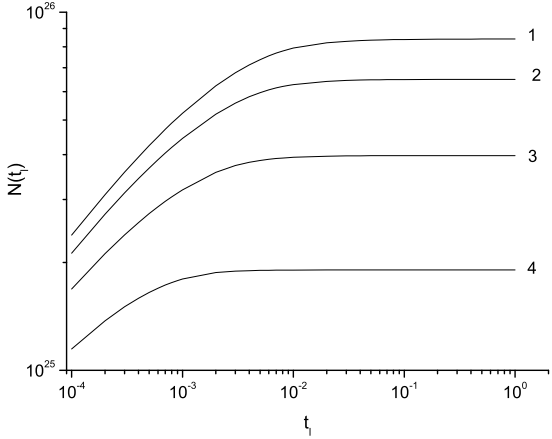}
\caption{ Total number of gamma-ray photons emitted by PBH that
can be detected by the Andyrchy array versus the time $t_l$ left
until the end of black hole evaporation: 1, $\theta = 0^\circ$; 2,
$\theta = 10^\circ$; 3, $\theta = 20^\circ$; 4, $\theta =
30^\circ$. } }
\end{figure}

\section*{Results}

We obtained a constraint on the number density of evaporating PBHs
using the technique described in Linton et al. (2006). Let a PBH
be located at distance $r$ from the array and be seen from it at
zenith angle $\theta$. The mean number of gamma-ray photons
detected by the array over the burst duration $t_b$ is then
$$\bar n(\theta) = \frac{\epsilon N(t_b(\theta))S(\theta)}{4\pi r^2}\; ,$$
where $S(\theta)$ is the array area and $\epsilon = 0.9$ is the
fraction of the events from a point source that fell into an
angular cell with a size of $7.0^\circ$. The number of bursts
detected over the total observation time $T$ can be represented as
$$N=\rho_{pbh} T V_{eff}\; ,$$
where
$$V_{eff}=\int d \Omega {\int \limits_{0}^{\infty} dr
r^2 F(n(\theta),\bar n(\theta), t_b(\theta))}$$ %
is the effective volume of the space surveyed by the array,
$\rho_{pbh}$ is the number density of evaporating PBHs, and $F(n,
\bar n, t_b)$ is the detection probability of a cluster of $n$
EASs with the mean value of $\bar n$ over the burst duration
$t_b$. In turn, the total detection probability can be expressed
as a product of the Poisson probability of $n$ EASs falling on the
array with the mean value of $\bar n$ and the probability to
detect all $n$ EASs over the burst duration $t_b$ with the array
dead time per event $t_d$:
$$F(n,\bar n,t_b) = f(n,t_b,t_d) \frac{e^{-\bar n} {\bar n}^n}{n!} \; .$$

\begin{figure}[!t]
{
\includegraphics[width=82mm]{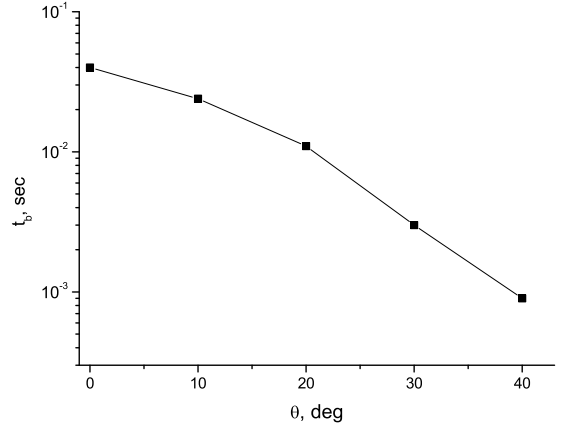}
\caption{ Burst duration versus zenith angle for the Andyrchy
array. } }
\end{figure}

In calculating the effective volume, we take $n(\theta) =
n'(\theta) + 1$ (this means that the distributions of the detected
clusters in multiplicity can be explained by the background of
chance coincidences). The probabilities $f(n, t_b, t_d)$ of
detecting $n(\theta)$ EASs over the burst duration $t_b(\theta)$
with the array dead time per event $t_d$ were calculated by the
Monte Carlo method by taking into account the burst time profile
for a given zenith angle. The effective volume of the space
surveyed by the array, $V_{eff}$, is $1.88 \times 10^{-9}$ pc$^3$,
with $N_0$ being $9 \times 10^{20}$. If the evaporating PBHs are
distributed uniformly in the local region of the Galaxy, then the
upper limit $\rho_{lim}$ on the number density of evaporating PBHs
at the $99\%$ confidence level can be calculated from the formula
$$\rho_{lim} = \frac{4.6}{V_{eff}\cdot T} \; ;$$
in our case, $\rho_{lim} = 8.2 \times 10^8$ pc$^{-3}$ yr$^{-1}$.

\section*{Comparison with the Results of Previous Studies and
Conclusions}

Previously, the high-energy GRBs from evaporating PBHs were
searched for in experiments on the CYGNUS (Alexandreas et al.
1993; the threshold energy of the primary gamma-ray photons is
$E_{th} \sim 30$ TeV), HEGRA (Funk et al. 1995; $E_{th} \sim 30$
TeV), and Tibet (Amenomori et al. 1995; $E_{th} \sim 10$ TeV) EAS
arrays for fixed one-second time intervals without taking into
account the burst duration (and its dependence on the zenith angle
for the specific array). In these experiments, constraints on the
number density of evaporating PBHs were obtained by assuming the
absence of any fluctuations in the number of EASs falling on the
array (it was assumed that $n = \bar n$) and by disregarding the
effects of the burst time profile and the array dead time on the
burst detection probability. The latter is particularly important
for arrays with a high gamma-ray detection threshold (i.e., with a
short burst duration $t_b$) and a relatively long dead time $t_d$.
Disregarding the effects of the array dead time and the burst time
profile on the burst detection probability when interpreting the
data will lead to an appreciable overestimation of this
probability (i.e., the factor $F$ in the above expression for the
effective volume $V_{eff}$) and, as a result, to an
underestimation of the upper limit on the PBH number density. For
example, in this case, a constraint of $3.1 \times 10^6$ pc$^{-3}$
yr$^{-1}$ instead of the real $8.2\times 10^8$ pc$^{-3}$ yr$^{-1}$
is obtained for the Andyrchy array. In the limiting case, the
array can be insensitive to the high-energy GRBs from PBHs
altogether, since for $t_d \ge t_b$ no more than one gamma-ray
photon can be detected over the burst duration and a burst of $n$
gamma-ray photons can be detected if $n t_d < t_b$.

\begin{figure}[!t]
{
\includegraphics[width=82mm]{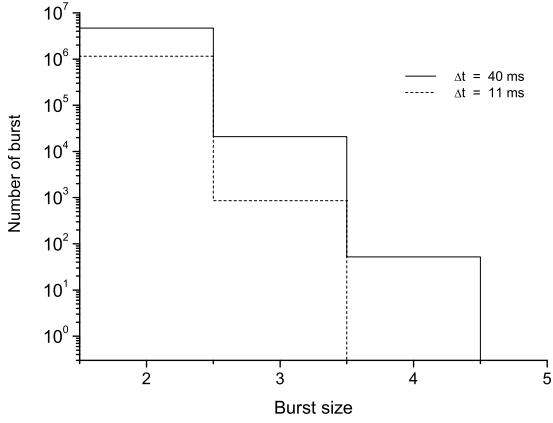}
\caption{ Measured number of clusters versus cluster size for two
time intervals, $\Delta t = 40$ ms and $\Delta t = 11$ ms. } }
\end{figure}

Figure 6a presents the results of the studies listed above as they
are given by their authors. This figure also shows the constraint
obtained at $E_\gamma \sim 1$ TeV in an experiment on the Whipple
Cherenkov telescope (Linton et al. 2006) for a time interval of 5
s. In addition, for comparison, it also shows our constraint that
we obtained in the approximation of a zero array dead time and by
disregarding the burst time profile. The effective energy of the
detected gamma-ray photons is along the horizontal axis. Two
constraints obtained for different event selection conditions are
given for the HEGRA array.

\begin{figure}[!t]
{
\includegraphics[width=82mm]{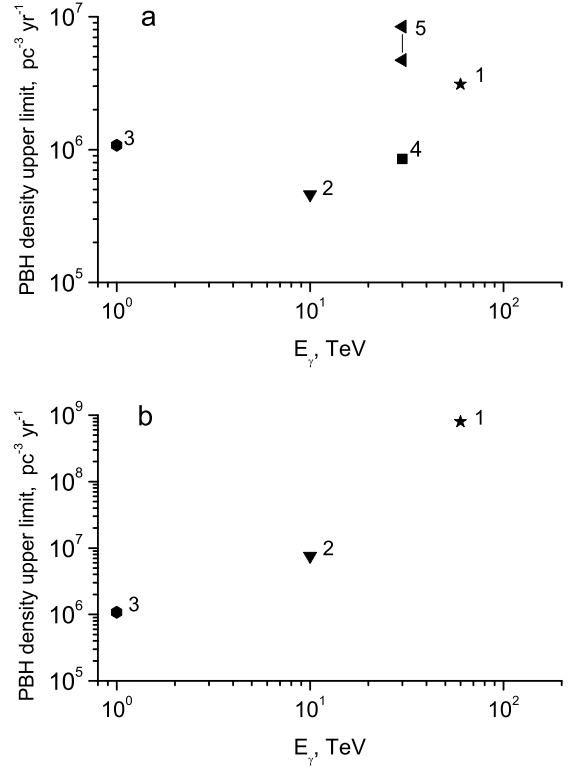}
\caption{ Upper limits on the number density of evaporating PBHs
for the evaporation model without a chromosphere versus effective
energy of the detected gamma-ray photons. Panel (a): 1, the
Andyrchy array (this paper, without the effects of the burst time
profile and the array dead time on the burst detection
probability); 2, original data from the Tibet array (Amenomori et
al. 1995); 3, the Whipple Cherenkov telescope (Linton et al.
2006); 4, original data from the CYGNUS array (Alexandreas et al.
1993); 5, original data from the HEGRA array (Funk et al. 1995).
Panel (b): 1, the Andyrchy array (this paper, with the effects of
the burst time profile and the array dead time on the burst
detection probability); 2, recalculation of the Tibet data (see
the text); 3, the Whipple Cherenkov telescope (Linton et al.
2006). } }
\end{figure}

The upper limits on the number density of evaporating PBHs with
allowance made for the burst time profile and the array dead time
are shown in Fig. 6b. We obtained the constraint for the Tibet
array ($7.6 \times 10^6$ pc$^{-3}$ yr$^{-1}$) by recalculating the
data from Amenomori et al. (1995) using the technique described
above (the dead time of this array is 5 ms per event). For the
CYGNUS array, the expected burst duration is no more than 40 ms
for a source at zenith, while the dead time estimated from the
data in Alexandreas et al. (1993) is $\sim 30$ ms per event. Since
these two times are of the same order of magnitude, the real
constraint on the PBH number density cannot be calculated. For the
HEGRA array, we also disregarded the effect of its dead time,
because no dead time is given in Funk et al. (1995). Finally,
estimates show that taking into account the dead time is
unimportant at relatively low gamma-ray photon energies ($\sim 1$
TeV). Therefore, the result of the Whipple array (Linton et al.
2006) remains essentially unchanged.

Comparison of Figs. 6a and 6b leads us to conclude that taking
into account the burst time profile and the array dead time is
very important at high energies of the detected gamma-ray photons.

How far are our constraints on the PBH number density from its
actual estimates? If we make two assumptions: (1) the spectrum of
the primordial density fluctuations is scale-invariant (in this
case, the data obtained in measurements of the relic gamma
radiation can be used to calculate the PBH mass spectrum and
number density) and (2) the PBH accumulation factor in the Galaxy
is equal to unity (i.e., there is no local increase in PBH number
density in the Galaxy), then the predicted number density of
evaporating PBHs is many orders of magnitude lower than the upper
limits shown in Fig. 6b. However, there are many cosmological
models in which the spectrum of the primordial density
fluctuations is nonmonotonic on small scales (see, e.g., Bugaev
and Klimai (2006) and references therein). The possible value of
the accumulation factor has already been discussed above. In a
word, we cannot rule out the possibility that the actual PBH
number density is close to the already available experimental
upper limits. In general, the predicted PBH number density in the
Universe is determined by our views of the end of inflation. PBHs
(even if they will not be discovered) are sources of unique
information about this period in the history of the early
Universe. Therefore, their searches will undoubtedly be continued.

As we see from Fig. 6b, the best (to date) constraint on the
number density of evaporating PBHs ($1.08 \times 10^6$ pc$^{-3}$
yr$^{-1}$) was obtained in the experiment on the Whipple Cherenkov
telescope (Linton et al. 2006). However, it should be noted that
the effective gamma-ray photon energy in our experiment is two
orders of magnitude higher than that in the Whipple one. Thus, our
upper limit pertains not to black holes in general, but to black
holes with certain properties (emitting 100-TeV gamma-ray photons
at the end of their evaporation during bursts lasting $\sim 10$
ms).

\section*{Acknowledgements}

This work was supported by the Russian Foundation for Basic
Research (project nos. 06-02-16135 and 07-02-90901), the "Neutrino
Physics" Basic Research Program of the Presidium of the Russian
Academy of Sciences, and the State Program for Support of Leading
Scientific Schools (project NSh-4580.2006.02).

\medskip
{\it \hfill Translated by V. Astakhov}

\end{document}